\documentclass[aps,
			   pre,
			   reprint,
			   showkeys, 
			   groupedaddress, 
			   onecolumn, 
			   secnumarabic, 
			   nofootinbib]{revtex4-2}

\usepackage{balance}
\usepackage{amsmath}
\usepackage{amssymb}
\usepackage{graphicx}

\usepackage{chr}

\newcommand{\ldis}{\ell_{\text{dis}}}

\hypersetup{
	pdfauthor={P. Dolai, S. Krekels and C. Maes},
	pdftitle={Inducing a bound state between active particles}
			}

\begin{document}
\title{Inducing a bound state between active particles}
\author{Pritha Dolai}
\author{Simon Krekels}\email{simon.krekels@kuleuven.be}
\author{Christian Maes}
\affiliation{Instituut voor Theoretische Fysica, KU Leuven}
\date{April 18, 2022}

\begin{abstract}
	We show that two active particles can form a bound state by coupling to a driven nonequilibrium environment.
	We specifically investigate the case of two \emph{mutually noninteracting} run-and-tumble probes moving on a ring, each in short-range interaction with driven colloids.  
	Under conditions of time-scale separation, these active probes become trapped in bound states.  
	In fact, the bound state appears at high enough persistence (low effective temperature).
	From the perspective of a co-moving frame, where colloids are in thermal equilibrium and the probes are active and driven, an appealing analogy appears with Cooper pairing, as electrons can be viewed as run-and-tumble particles in a pilot-wave picture. 
\end{abstract}
\maketitle
\section{Introduction}
Bound states are ubiquitous in physics, from the atomic nucleus to planetary systems and galaxies. 
Such states appear when the interactions tend to localize particles to a more-or-less fixed distance from each other, stable under small perturbations.
Usually the interactions causing such behavior are attractive.
Bound states are more exceptional when the (only) direct interaction between particles is repulsive.  
A famous example is that of the Cooper pair, where electrons repel each other through the Coulomb interaction, but form a pair in a low-temperature phonon bath \cite{cooper1956bound,bardeen1957microscopic,bcs1957theory}.
The electron-phonon interactions indeed induce a  small attraction between the electrons leading to a paired state of electrons with lower energy than the Fermi energy.

Active particles have recently gathered widespread interest as a biophysically inspired example of far-from equilibrium systems \cite{ramaswamy2017active, gompper2020motile, romanczuk2012brownian, marchetti2013hydro, bechinger2016environment}.
The local dissipation of energy and generation of motility not only gives rise to emergent phenomena at large scales like non-equilibrium ordering transitions and pattern formation but also exhibits many interesting features even at the single particle level \cite{malakar2018steady}. 
Run-and-tumble particles (RTPs) are the simplest realization of scalar active systems \cite{tailleur2008statmech}.  
A pair of interacting run and tumble particles have been shown to form bound states \cite{jamming,ledoussal,das2020gapstat,abhishek, mallmin2019exact} and represent the few particle versions of clustering as observed, for example, in motility-induced phase separation \cite{mips,redner2013structure,reichhardt2014jamming}.

The present paper reports on a similar effect: the emergence of a bound state, but between \emph{independent} (i.e., non-interacting) probes.  
More precisely, coupling of probes to a driven nonequilibrium medium induces an effective repulsive interaction  \cite{maes2017nonreactive, moscow}.
This repulsion can be counteracted by letting the probes become active, i.e., adding persistence over time-scales exceeding the relaxation time. This causes the probes to move towards each other, and in combination with the induced repulsion, cause the binding.
In other words, the repulsion is induced by contact with the nonequilibrium medium while the attraction is caused by adding activity, both effective.

In Section \ref{sec:setup} we present the details of our main example: active probes moving in a bath of driven colloids, trapped on a ring.  Section \ref{res} summarizes the main results.  
In Section \ref{coo} we expand on the analogy with the emergence of Cooper pairs of electrons. 
There the probes (electrons) are driven while the colloids form an equilibrium particle-bath.
That situation is, however, equivalent to the case with driven colloids in a moving reference frame.

\section{Setup}\label{sec:setup}

\begin{figure}
	\centering
	\includegraphics[width=.4\linewidth]{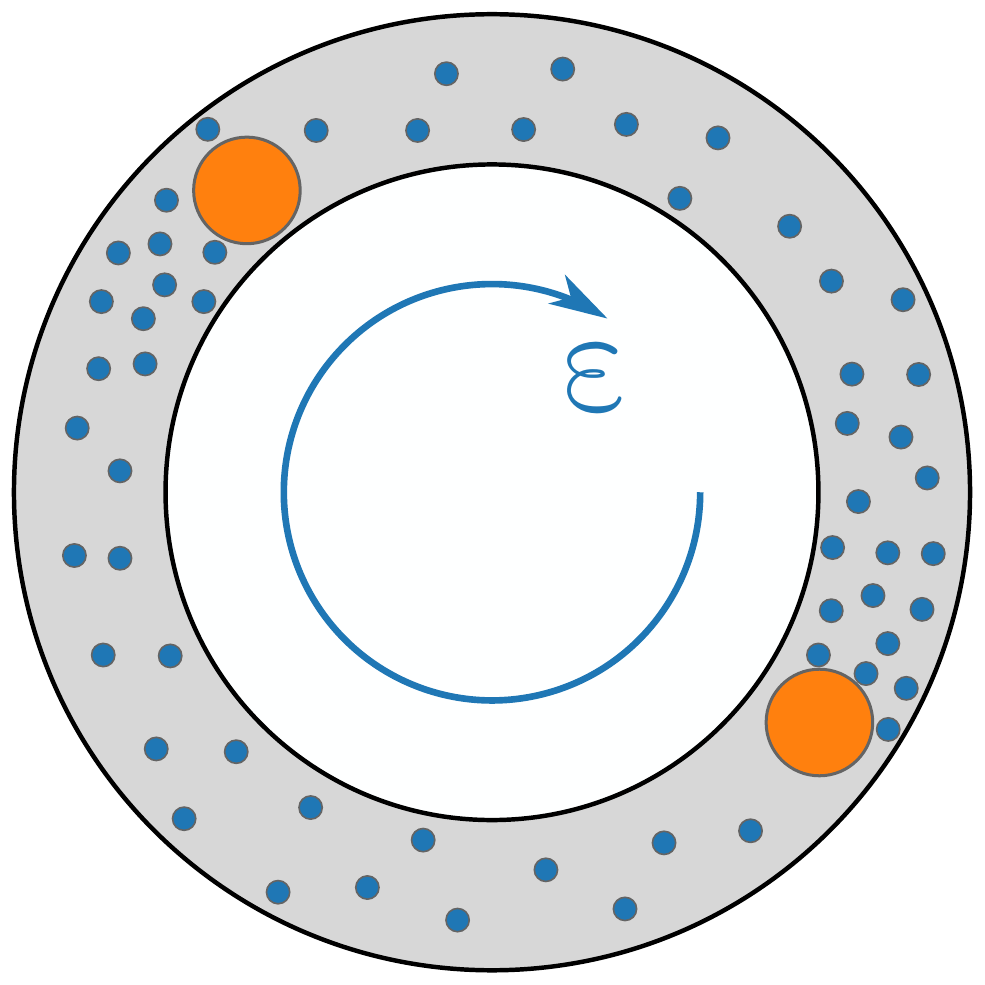}
	\caption{A schematic view of our model. The gray background represents the viscous thermal bath; the blue colloids are under global driving $ \varepsilon $ while the orange probes only interact with the colloids through a local interaction. The mathematical description of the model is given by eqs. \eqref{eq:eom_model_active}.}
	\label{fig:schematic}
\end{figure}

We start by explaining the mathematical ingredients of the model.  
Physically, we imagine two slow-moving, active probes colliding with fast, driven colloids, all confined in a toroidal trap and suspended in a viscous thermal bath.  
The global driving of the colloids can be optical (as e.g. in \cite{ciliberto}) or an electromagnetic force when the colloids are charged. In our model, the toroidal trap is a one-dimensional ring which is simpler to analyze mathematically.
We model the probes as interacting through soft-core repulsion to allow the colloids to pass through the probes as they would be able to move past the probes in a two-dimensional toroidal geometry.
We checked by simulation that at least in the quasi-static case (cf. Section \ref{sec:pass_QS}) the same phenomenology can be reproduced in two dimensions on a ring-shape, compare Fig. \ref{fig:schematic}.

\subsection{The Model}

The model under consideration consists of two main components: a ``medium'' of $N$ fast-moving colloids and two slow-moving probes. 
These probes and colloids move on a ring of length $L$, that is: a one-dimensional line with periodic boundary conditions.  
The colloids are subject to friction and noise from a thermal environment (represented by temperature and friction parameters).  
The colloids are globally driven by a constant external force $ \varepsilon $, and are independent, i.e., mutually non-interacting. 
The only explicit interaction is between probes and colloids through a local potential $ u(z) $.  
The probes are active, and subject to run-and-tumble dynamics.  
The whole system can be described by the overdamped equations of motion
\begin{equation}\label{eq:eom_model_active}
	\begin{aligned}
		\zeta \dot{\eta}_i(t) &= \varepsilon - \sum_{\alpha=1,2}u'\big(x_{\alpha}(t) - \eta_i(t)\big) + \left (\frac{2\zeta}{\beta}\,\right )^{\frac{1}{2}}\,\xi_i(t)\\
		\gamma \dot{x}_{\alpha}(t) &= \nu_{0}\sigma_{\alpha}(t) -\sum_{i=1}^Nu'\big(\eta_i(t) - x_{\alpha}(t)\big)
	\end{aligned}
\end{equation}
The  $ \eta_i(t) $ represent the positions of the $i$-th colloid; for  $\alpha=1,2$ the $x_1(t),x_2(t)$ represent the positions of the active probes at time $t$.  
The $ \zeta $ and $ \gamma $ are friction coefficients for the colloids and probes respectively, and $\xi$ is the standard white noise representing, in an effective way, the presence of a thermal viscous environment at inverse temperature $\beta$.
The probes are not subject to thermal noise, as they are thought of as being objects of a larger length scale than the colloids, but to dichotomous noise $\sigma_{\alpha} = \pm 1$  with flipping rate $ \tau^{-1} $. The $ \nu_0 $ represents the strength of the active force. 
The probe-colloid interaction potential $u(z)$ is taken to be
\begin{equation}\label{eq:sim_potential}
u(z) = \left\{\begin{array}{ll}
u_0\left(1-\left(\dfrac{z}{\delta}\right)^2\right)^2, & |z|<\delta\\
0\,, & |z|>\delta
\end{array}\right.
\end{equation}
The parameter $u_0$ governs the strength of the potential and the range $\delta$ can be thought of as the radius of the probes.  
Since the potential is finite, reaching a maximum at $z=0$, the $ u(z) $ represents a soft-core repulsive interaction.  
The precise shape of the potential is not very important, see \cite{maes2017nonreactive}; here we take the potential to be repulsive ($ u_{0}>0 $), but attractive potentials also give rise to the same phenomenology, as is demonstrated in \cite{maes2017nonreactive}.

\begin{figure}
	\centering
	\includegraphics[width=.6\linewidth]{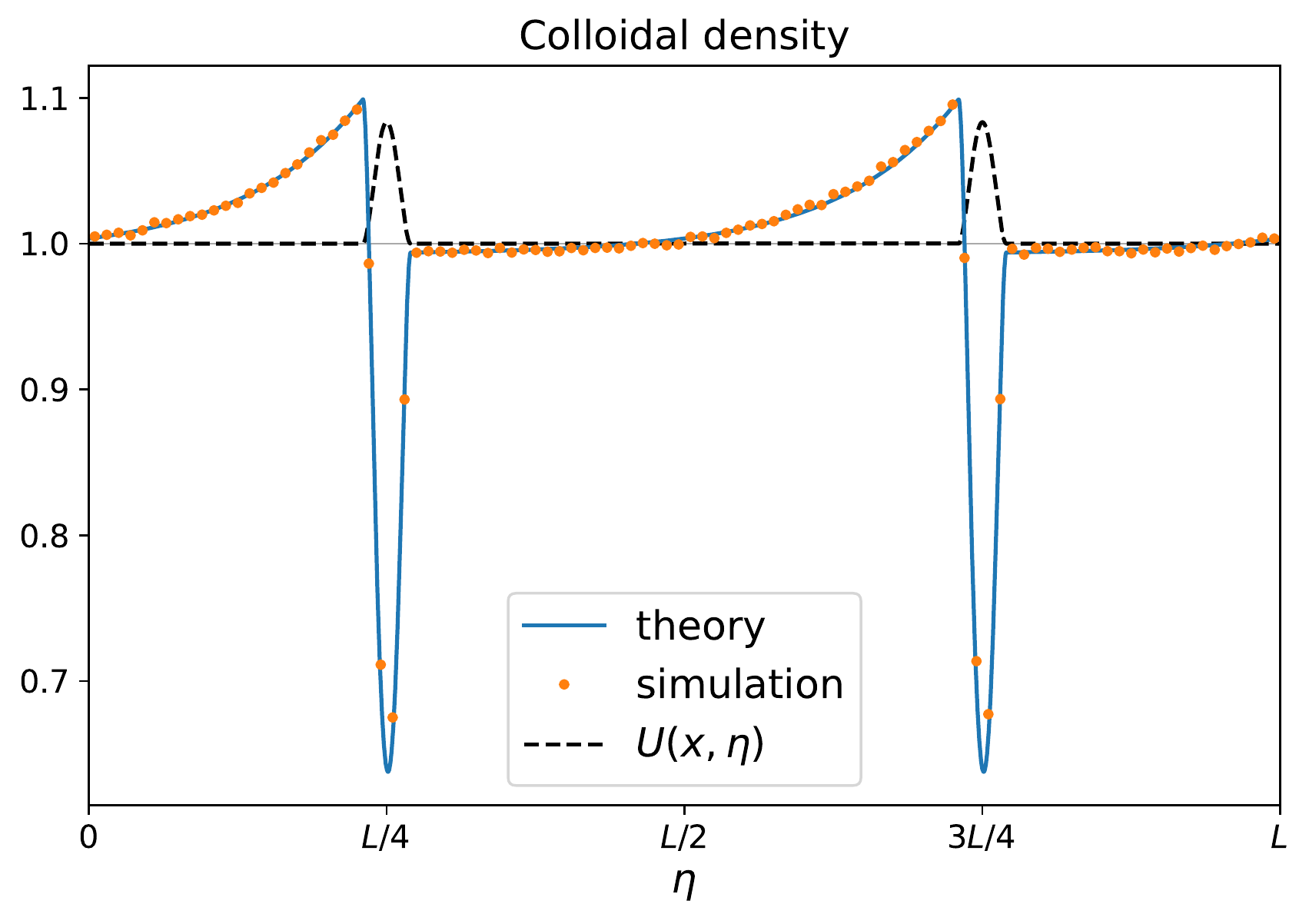}	
	\caption{Colloidal density profile. Comparison of theory \eqref{app:dens} and simulation for the stationary colloid distribution with tails induced by fixed probes at positions $ x_{\alpha} $. The density is expressed relative to the mean density $ N/L $. The coordinate $\eta$ represents the position of the colloids. In this figure, $ \delta = L/50 $ and $ \ldis=L/10 $. The probes are held fixed at $x_\alpha = (L/4, 3L/4)$ while the colloids are driven by $\varepsilon=2$. }
	\label{fig:tails}
\end{figure}

\subsection{Passive case, quasi-static limit}\label{sec:pass_QS}

Except for the active motion of the probes, the model \eqref{eq:eom_model_active} has been treated before in  \cite{maes2017nonreactive,moscow}.   
The colloidal driving introduces a \emph{dissipative length scale} $\ell_\text{dis} =  (\varepsilon\beta)^{-1}$.  The most interesting nonequilibrium effects appear when $2\delta\lesssim\ell_\text{dis} \ll L$:  the work done by the driving in one cycle ($L$) must be higher than the strength of thermal fluctuations, while those thermal fluctuations must on the other hand be big enough to break possible bonding ($\delta$). 
It was shown in \cite{maes2017nonreactive}  that the mere presence of the probes (kept fixed in time at some $x_\alpha$) generates exponentially decaying tails of characteristic length $ \ell_\text{dis} $ in the stationary colloid density profile $\rho_{x}(\eta)$.
These results are summarized in Appendix \ref{sec:coll_profile}.
These tails share their origin with the exponential density profile obtained in the context of sedimentation of colloids \cite{sedimentation} (in that case $\varepsilon = mg$).
The derivation of the shape of these tails was done in the \emph{quasi-static} limit, where the relaxation time of the colloids is much smaller than that of the probes, i.e., the positions of the probes change over a far greater time scale than those of the colloids. Hence, when considering the stationary density profile of the colloids, the probes can be assumed to remain fixed as compared to the colloids. 
In simulations (see Appendix \ref{app:simulations} and \cite{gitlab}) we retain the time scale separation by letting $ \gamma\gg\zeta $.

The tails can be reproduced by simulating eqs. \eqref{eq:eom_model_active}, keeping the probes fixed (i.e., $ \gamma \to \infty $). 
The result and comparison with the predicted profile are shown in Fig.~\ref{fig:tails}.

For our purposes it is important that these long tails induce a repulsive interaction between the (quasi-static) probes. 
In the quasi-static limit the colloid density reaches a stationary  distribution $\rho_x$, with $x$ representing the (static) probe positions (see Appendix \ref{sec:coll_profile}), and the induced force on the $\alpha$-th probe is
\begin{equation}\label{eq:avg_force}
f_\alpha(x) = - \int_{0}^{L} u'(x_\alpha - \eta)\,\rho_x(\eta)\id \eta
\end{equation}
This force is calculated explicitly in \cite{maes2017nonreactive} and can be decomposed as
\begin{equation}\label{eq:f_int}
f_{\alpha} = f^{\text{drift}} + f_{\alpha}^{\text{int}}
\end{equation}
Here, $f^{\text{drift}}$ is proportional to the colloidal current through the ring and is equal for both probes. 
The interaction terms $f_\alpha^{\text{int}}$ are induced solely by the tails in the colloid distribution, and decay exponentially in the distance between the probes,
\begin{equation}\label{eq:f_int_expl}
f^{\text{int}}_{\alpha}(x_{1},x_{2}) \propto e^{-(x_1 - x_2)^+/\ell_\text{dis}}
\end{equation}
where $ (x_1 - x_2)^+ $ signifies the one-way periodic separation,
\begin{equation}\label{eq:onewaysep}
(x_1 - x_2)^+ = \left\{\begin{array}{lcl}
x_1 - x_2 & \text{if} &x_1 > x_2 \\
L - (x_2 - x_1) & \text{if} & x_1 < x_2
\end{array}\right .
\end{equation}
Thus, $ f^{\mathrm{int}}_{\alpha} $ is not symmetric around $ x_{\alpha} $, but one-sided (cf. Fig. \ref{fig:tails}: the colloid tails are asymmetric, leading $ f^{\mathrm{int}}_{\alpha} $ to also be asymmetric).
In the limit $ \ell_\text{dis}\ll L $,  the range of the repulsion is small compared to the size of the ring, and the probes behave approximately as interacting hard spheres. 
In any cases the probes obtain an effective mutual repulsion induced by the colloids.
Indeed, in the quasi-static limit there is a force 
\begin{equation}\label{fd}
F(d) = f^{\text{int}}_{1} - f^{\text{int}}_{2} \propto e^{-\frac{d}{\ell_\text{dis}}} - e^{\frac{d-L}{\ell_\text{dis}}}
\end{equation}
dependent only on $d = (x_{1}-x_{2})^{+}$, the separation between the two probes, and which is zero at $d = L/2$. Note that $ d $ is not the smallest periodic distance between the two probes, but rather a measure of how far \emph{ahead} probe 1 is of probe 2. 
Thus, $ d $ runs from $ 0 $ to $ L $. Swapping the probe labels $ x_{1} $ \& $ x_{2} $ has the effect of mirroring $ d $ around $ L/2 $.

In the quasi-static limit, the probe separation dynamics for the passive ($ \nu_{0}=0 $) case reduces to
\begin{equation}\label{eq:dyn_d}
\gamma\dot{d} = \kappa\left(e^{-\frac{d}{\ell_\text{dis}}} - e^{\frac{d-L}{\ell_\text{dis}}}\right)
\end{equation}
where $ \kappa $ is a coefficient characterizing the strength of the force, and can be calculated from knowledge of the stationary profile $\rho_{x}$, see \eqref{eq:kappa}. The analytic solution to \eqref{eq:dyn_d}, derived in Appendix \ref{sec:sol_d}, is
\begin{equation}\label{eq:sol_d}
d(t) = \frac{L}{2}-\ell_\text{dis} \log\frac{C(t)-1}{C(t)+1}
\end{equation}
where
\begin{equation*}
C(t) = \left(\frac{e^{-d_{0}/\ell_\text{dis}}+\sqrt{\Lambda}}{e^{-d_{0}/\ell_\text{dis}}-\sqrt{\Lambda}}\right)e^{\frac{2\kappa\sqrt{\Lambda}}{\ell_\text{dis}\gamma}t},\qquad \Lambda = e^{-\frac{L}{\ell_\text{dis}}}
\end{equation*}
with $ d_{0} $ the initial separation.
This solution gives a sense of the time scale of relaxation of the probes towards $ L/2 $, 
\begin{equation}
t_{\mathrm{rel}}^{0} = \frac{\ell_\text{dis}\gamma}{2\kappa}\,e^{\frac{L}{2\ell_\text{dis}}}
\end{equation}
Notice that $ t_{\mathrm{rel}}^{0} $ depends strongly on the size of the ring; the characteristic time increases exponentially with $ L/\ldis $, becoming very large in the limit $ \ldis\ll L $. 
In that limit, the probe interactions approximate those of hard spheres, where outside of the immediate interaction range, $ f^{\mathrm{int}}_{\alpha}\approx 0 $, and therefore the relaxation time to $ L/2 $ diverges.

\subsection{Quasi-static motion of active probes}\label{sec:QS}

\begin{figure}
	\centering
	\includegraphics[width=.6\linewidth]{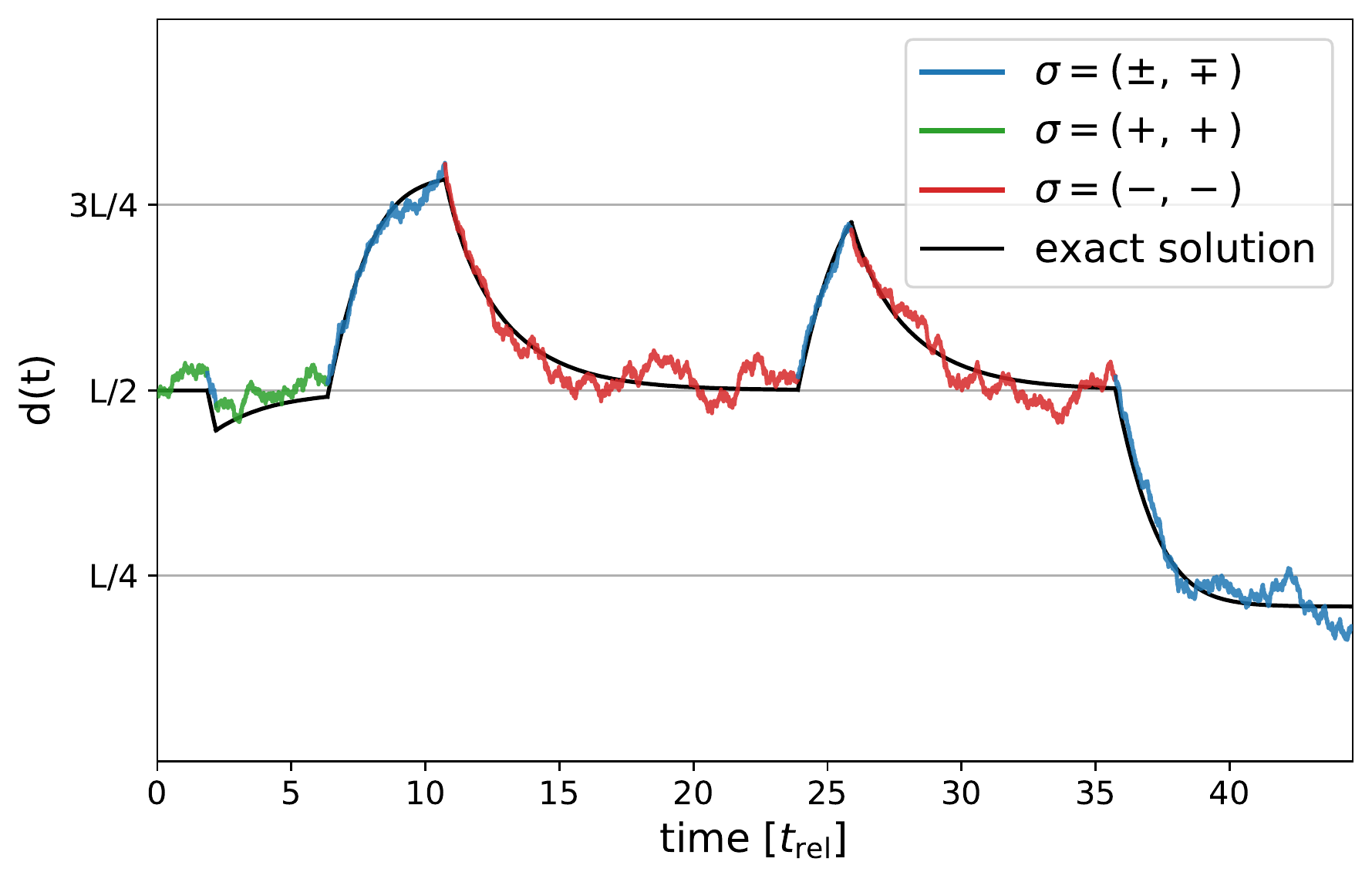}
	\caption{Simulated trajectory of the separation of two run-and-tumble probes in a driven colloid bath. The colors represent the mutual state of the probes, and the black line represents the exact solution to \eqref{eq:eom_d_active}. The parameters used are in Tab. \ref{tab:params}, except $\gamma = 5\times 10^{3}$ and $ \tau = 6.7\,t_{\mathrm{rel}} $. Note that the exact solution was obtained in the quasi-static limit while the simulated trajectories are governed by \eqref{qsdd} and are thus subject to colloidal noise.}
	\label{fig:tumbles_colorcoded_theory}
\end{figure}

We re-introduce the active force into \eqref{eq:dyn_d} to get the separation dynamics for self-propelling probes
\begin{equation}\label{eq:eom_d_active}
\gamma \dot{d} = \nu_{0}[\sigma_{1} - \sigma_{2}] + F(d)
\end{equation}
where $ F(d) $ is given by \eqref{fd}.  
The noise  $\sigma_{1} - \sigma_{2}$ can take three values.
Eq. \eqref{eq:eom_d_active} describes a run-and-tumble process where the noise is no longer dichotomous, but three-valued \cite{santra2020rtp2d}.
Such systems have been considered before, subject to a harmonic confining potential \cite{urna}.  
Our potential is exponential, as may be verified by integrating $ F(d) $,
\begin{equation*}
	V(y) = -\int_{L/2}^{y} F(z)\, \dif z = \kappa\ldis\left(e^{-y/\ldis}+e^{(y-L)/\ldis}\right)
\end{equation*}
up to a constant.

Fixing any one of the three states $ \sigma_{1}-\sigma_{2} = \{-2, 0, +2\} $, an exact solution to \eqref{eq:eom_d_active} is derived in Appendix \ref{sec:sol_d}.  Putting (and fixing) 
$\nu_{0}[\sigma_{1} - \sigma_{2}] =\nu$, the solution to \eqref{eq:eom_d_active} is
\begin{equation}\label{qsa}
d(t) = -\ldis\log\left[-\frac{\mu}{2\kappa}\left(\frac{1+C(t)}{1-C(t)}\right)-\frac{\nu}{2\kappa}\right]
\end{equation}
where
\begin{equation*}
	C(t)=\left(\frac{2\kappa e^{-\frac{d_{0}}{\ldis}}+\nu+\mu}{2\kappa e^{-\frac{d_{0}}{\ldis}}+\nu-\mu}\right)e^{\frac{\mu}{\gamma\ldis}t}, \quad \mu = \sqrt{\nu^{2}+4\Lambda\kappa^{2}}
\end{equation*} 
The characteristic relaxation time is
\begin{equation}\label{eq:trel}
t_{\text{rel}} = \frac{\ldis\gamma}{2\sqrt{\nu^2+\Lambda\kappa^2}}
\end{equation} 
The density plot will be determined by the ratio of the lifetime $ \tau $ to this characteristic time $ t_{\mathrm{rel}} $. 
Note that for different values of $ \sigma_{1}-\sigma_{2} $, there are different solutions \eqref{qsa} and therefore also different timescales \eqref{eq:trel}. 
The relaxation towards the state with $\nu=0$ generally takes longer than the relaxation to any of the states with $ \nu \neq 0 $.
A sample trajectory from simulation together with the solution \eqref{qsa} is shown in Fig. \ref{fig:tumbles_colorcoded_theory}.
The relative state of the two probes, i.e., the state of the three-state RTP, is marked by the color-coding of the simulated trajectory. 

\section{Results}\label{res}

\begin{figure}
	\centering
	\includegraphics[width=.6\linewidth]{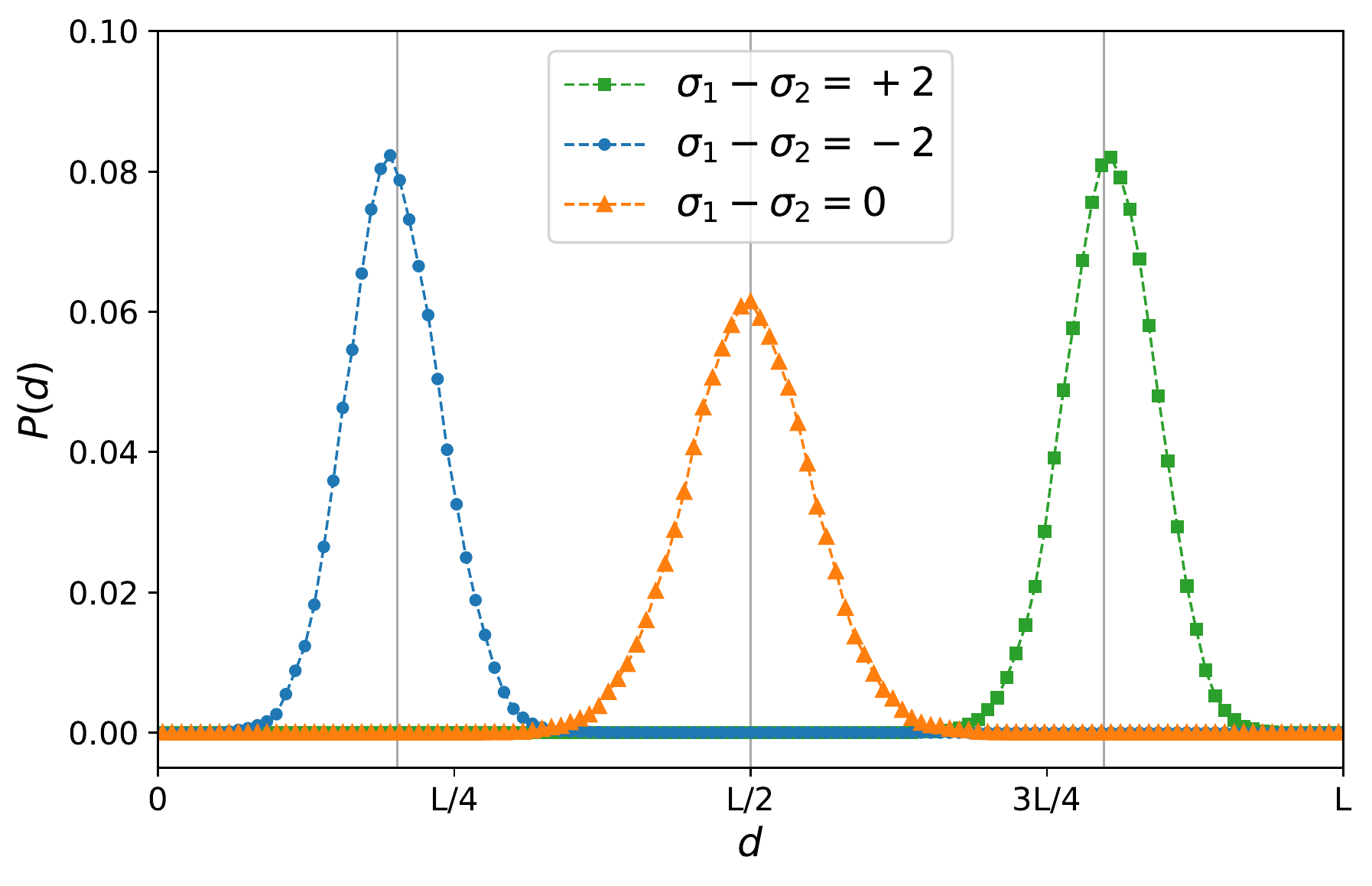}
	\caption{Density plots representing the distribution of separations of the probes for different relative velocities $ (\sigma_{1}-\sigma_{2}) $. The case where $ (\sigma_{1}-\sigma_{2}) = 0$ represents the ``passive'' case, while the case $ (\sigma_{1}-\sigma_{2}) = -2$ represents the bound state. The simulation parameters used are listed in Tab. \ref{tab:params}. The case $ (\sigma_{1}-\sigma_{2})=+2 $ was simulated with the \emph{probes} being driven instead of the colloids being driven (cf. \eqref{eq:moving_frame}). The vertical gray lines are the $ d^{*}_{\pm} $.}
	\label{fig:histogram_passive}
\end{figure}
\begin{figure*}
	\centering
	\includegraphics[width=\linewidth]{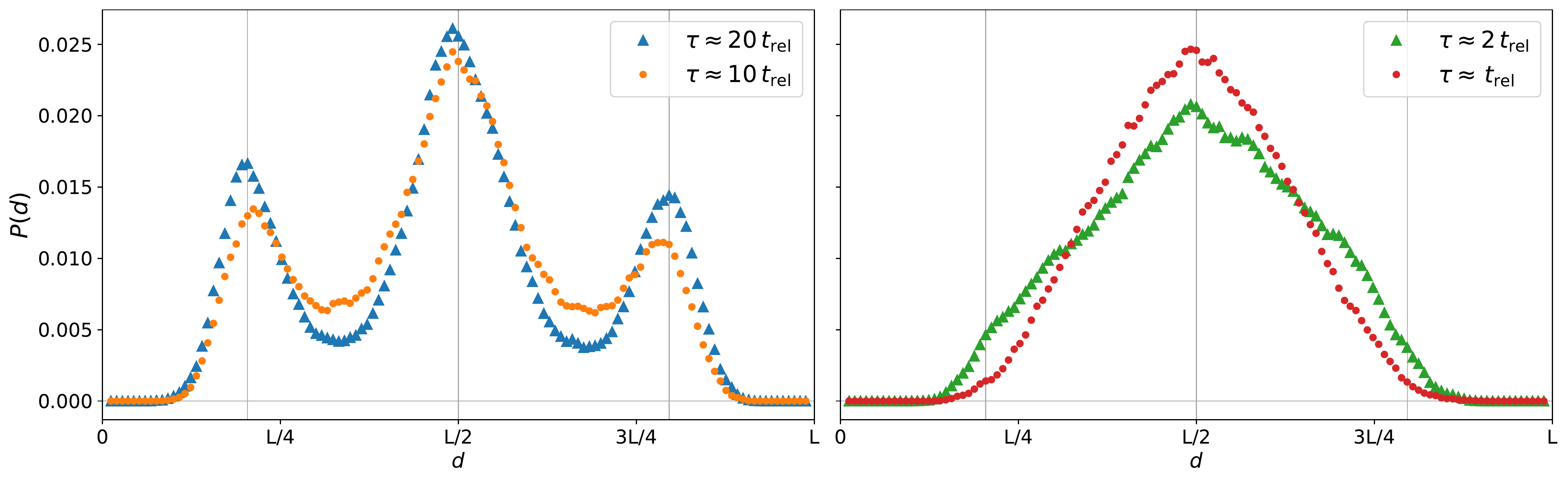}
	\caption{Density plots representing the distribution of the probe separations for different tumbling rates. The distributions with higher persistence have a clear three-peak structure, while those with lower persistence have a single central peak. The gray vertical lines show $ d^{*}_{\pm} $. The simulation parameters used are listed in Tab. \ref{tab:params}. Note the symmetry around $ L/2 $. This follows from the definition of $ d $, cf. \eqref{eq:onewaysep}. If we restrain from labeling the probes, the bound states left and right of $ L/2 $ are physically identical.}
	\label{fig:shape_trans}
\end{figure*}

Up to this point, we have worked in the quasi-static limit, which must be abandoned when turning to simulations with moving probes.
It is fair, however, to expect that the bound state survives beyond the quasi-static limit, where the dynamics of the separation between the probes follow simply from \eqref{eq:eom_model_active}, 
\begin{equation}\label{qsdd}
\gamma \dot{d} = \nu_{0}[\sigma_{1} - \sigma_{2}] -\sum_{i=1}^{N}\big(u'(x_1-\eta_{i}) - u'(x_2-\eta_{i})\big)
\end{equation}
and hence is subject to the colloidal noise.
We now deal with fluctuations in the colloid profile, and the active force $ \nu $ is not constant, but follows a three-state run-and-tumble process with tumble rate $ 2\tau^{-1} $.  
The active noise may cause modifications to the tail and thus the repulsion (by modifying $ \ell_\text{dis} $). However, this effect is negligible as long as $\gamma\gg\zeta$. 

What happens is as follows: the active noise allows the probes to move towards each other ($ \sigma_{1}-\sigma_{2} \neq 0 $), counteracting the induced repulsion and thus forming a \emph{bound state}, where the probes stay at a more-or-less fixed distance from each other. 
Of course the lifetime of the bound state is limited by the flipping rate of $ \sigma_{ \alpha} $, and the system must be able to relax to its bound state before $ \sigma_{ \alpha} $ flips, that is, $ \tau > t_{\mathrm{rel}} $.

The obtained solution \eqref{qsa} in the quasi-static case is compared against simulations of \eqref{qsdd} in Fig.~\ref{fig:tumbles_colorcoded_theory}.
This confirms that the quasi-static approximation still holds for moving probes.
The parameters in the simulations are listed in Tab. \ref{tab:params} and are always used unless mentioned otherwise.
\begin{table}[h]
	\caption{The values used for simulations unless noted otherwise.}
	\label{tab:params}
	\setlength\tabcolsep{.5em}
	\begin{tabular}{c|ccccccccc}
		parameter & $ \delta $ & $ L $ & $ \varepsilon $ & $ \zeta $ & $\gamma$ & $ T $ & $\nu_{0}$ & $ \rho_0 $ & $ u_{0} $\\
		\hline
		value & $ \frac{1}{2} $ & 25 & 1 & 1 & $ 10^{3} $ & 5 & $ \frac{1}{2} $ & 200 & 2.5
	\end{tabular}
\end{table}

The bound state may be observed by recording the separations of the probes and plotting the histogram of these observed separations. 
Such a result is shown in Fig.~\ref{fig:histogram_passive}.
The separation at which the bound state settles can be calculated by setting 
\begin{equation}\label{eq:force_prop}
	F_{\nu}(d^*) = \kappa\left(e^{-\frac{d}{\ell_\text{dis}}} - e^{\frac{d-L}{\ell_\text{dis}}}\right) + \nu = 0
\end{equation} 
and extracting $d^*$ as,
\begin{equation*}
d^* = -\ldis\log \frac{\sqrt{\nu^2+4\Lambda\kappa^2}-\nu}{2\kappa}.
\end{equation*}

While the passive ($ \sigma_{1}-\sigma_{2} = 0 $) distribution, $ P(d) $, is symmetric around $ L/2 $, the propulsive distribution is not symmetric around $ d^* $ (see Fig. \ref{fig:histogram_passive}). This is because the linear approximation around the fixed point $ d^* $ does not suffice as an approximation. 
Expanding $ F_{\nu}(d) $ around $d^{*}$, 
\begin{equation}\label{eq:F_expansion}
F_{\nu}(d^{*}+\epsilon) = -\dfrac{\sqrt{\nu^{2}+4\Lambda\kappa^{2}}}{\ldis}\epsilon - \frac{\nu}{2\ldis^{2}}\epsilon^{2}+\mathcal{O}\left(\epsilon^{3}\right)
\end{equation}
which properly captures the asymmetry in the bound state density plot. 
The first order term can be interpreted as the spring constant in the linearized system, and although it is not an adequate approximation, it can still offer some insight.
When noise is added to the system through fluctuations in the colloid profile, this first term is the main restoring force, and this determines the width of the stationary distribution around $ d^{*} $. Thus, the stronger the propulsion the narrower the peak will be. 
Also, smaller $\ldis$ leads to narrower peaks.

Depending on the state the two probes are in, the distribution of their separations tends towards either a bound state (the distribution for $ (\sigma_{1}-\sigma_{2})=-2 $ in Fig~\ref{fig:histogram_passive}, or its equivalent mirrored around $ L/2 $) or a repulsive state ($ (\sigma_{1}-\sigma_{2}) = 0 $ in Fig.~\ref{fig:histogram_passive}).

If, however, the rate is chosen larger than the inverse of the relaxation time, $ \tau < t_{\mathrm{rel}} $, the system generally does not relax to the bound state, and the peaked structure of the separation distribution is lost to a distribution more uniformly peaked around $L/2$. 
This is the so-called \emph{shape transition} covered in \cite{urna, abhishek} for a noiseless system in a harmonic potential.
The shape transition can be reproduced for the model considered here by simulating the system with varying tumble rates. 
The result is shown in Fig.~\ref{fig:shape_trans}.
Note that for lower persistencies (smaller $ \tau $) the peaks at the edges, which represent the bound states, disappear.

\begin{figure}
	\centering
	\includegraphics[width=.6\linewidth]{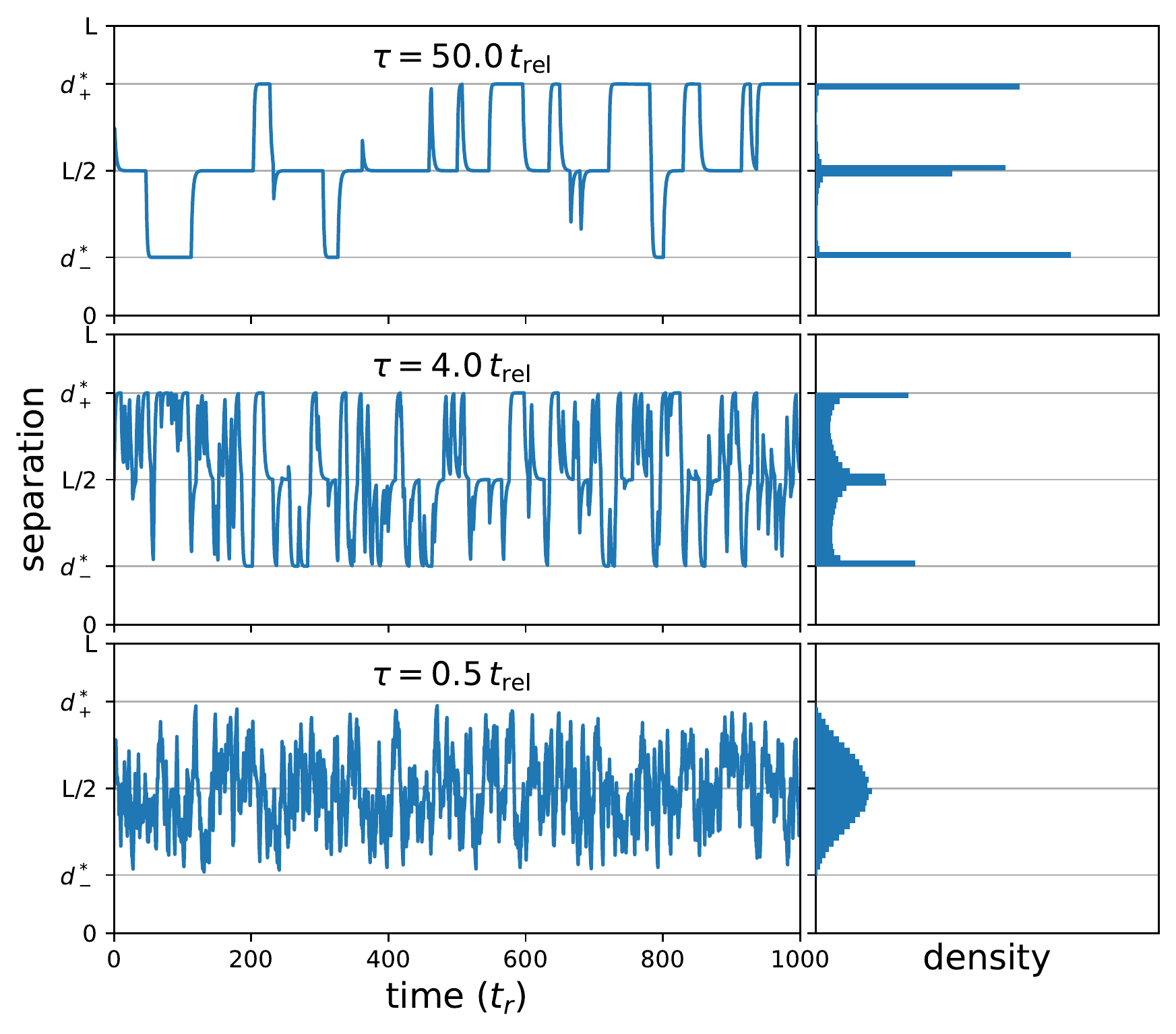}
	\caption{Trajectories for the probe separation calculated by using \eqref{qsa}. All parameters were set to those used in simulations (Tab. \ref{tab:params}).}
	\label{fig:sep_dynamics}
\end{figure}

The dynamical origin of this transition can easily be visualized.
Three trajectories of the probe separation dynamics, each with a different tumble rate are shown in Fig. \ref{fig:sep_dynamics}.
These trajectories were calculated using the exact solution for the quasi-static case \eqref{qsa}.
It clearly shows why no bound state peaks can form for higher tumble rates; the probes do not reach a bound state before they tumble again.

At constant persistence $ \tau $, the shape transition can be triggered by tweaking $ t_{\mathrm{rel}} $ so that $ \tau < t_{\mathrm{rel}} $ for $ \nu \neq 0 $. 
One obvious way to achieve this is by increasing the value of $ \ldis $, which is equivalent to increasing the temperature.
All other parameters kept fixed, the relaxation time $ t_{\mathrm{rel}} $ \eqref{eq:trel} can be seen to increase monotonically with $ \ldis $ and thus we conclude that bound states form at low temperature.

\section{Analogy with Cooper pairing}\label{coo}
Thus far, we have observed pairing effects between active (run-and-tumble) probes in a passive, yet driven bath. 
The repulsive interaction which together with the active propulsion enables the bound states to form, is entirely induced by the driven bath.
This phenomenology is reminiscent of the way Cooper pairs are formed when electrons obtain an effective attractive interaction induced by the phonon bath.
The similarity becomes even more striking when the electrons are treated in a pilot-wave picture, since then they have properties which are very similar to those of run-and-tumble particles.

We briefly explain, see \cite{ward,dirac} for more details.  A  Dirac  electron is described by a bispinor
\begin{equation*}
	\psi = \begin{pmatrix}
	\phi_{+}\\\phi_{-}
	\end{pmatrix}
\end{equation*}
where $ \phi_{\pm} $ are each themselves two-component spinors. 
The bispinor $ \psi $ obeys the Dirac equation\footnote{In this section, we will use natural units: $ \hbar = c = 1 $.},
\begin{equation}\label{eq:dirac}
	\imu\gamma^{\mu}\partial_{\mu}\psi - m\psi = 0
\end{equation}
with associated current
\begin{equation}\label{eq:dirac_current}
	j_{D}^{\mu} = \bar{\psi}\gamma^{\mu}\psi
\end{equation}
where the $ \gamma^{\mu} $ are the gamma-matrices and $ \bar{\psi} = \psi^{\dagger}\gamma^{0} $. Here, we work in the Weyl representation of the $ \gamma $-matrices, but the results hold for arbitrary representations \cite{ward}. 

The four-component Dirac spinor can be decomposed into left- and right-handed chiral components using the projection operators $ P_{\pm} = (1\pm\gamma_{5})/2 $ such that
\begin{equation*}
	\psi_{+} = P_{+}\psi = \begin{pmatrix}
	\phi_{+}\\0
	\end{pmatrix} \qquad \psi_{-} = P_{-}\psi = \begin{pmatrix}
	0\\\phi_{-}
	\end{pmatrix} 
\end{equation*}
The Dirac equation then decomposes into two equations
\begin{equation}\label{eq:chiral_diraceq}
	\imu\gamma^{\mu}\partial_{\mu}\psi_{\pm} - m\psi_{\mp} = 0
\end{equation}
and the Dirac current decomposes as $ j_{D}^{\mu} = j_{+}^{\mu} + j_{-}^{\mu} $, with
\begin{equation*}
	j_{\pm}^{\mu} = \bar{\psi}_{\pm}\gamma^{\mu}\psi_{\pm}
\end{equation*}
Unlike $ j_{D}^{\mu} $, the $ j_{\pm}^{\mu} $ are not individually conserved, but their continuity equations satisfy
\begin{equation}\label{eq:chiral_continuity}
	\partial_{\mu}j_{\pm}^{\mu} = \pm F, \hspace{.1\linewidth} F = 2m\,\Im\left(\psi_{+}^{\dagger}\gamma^{0}\psi_{-}\right)
\end{equation}

these equations can be written more explicitly as

	\begin{equation}\label{inr}
		\frac{\partial \rho_{\pm}}{\partial t}(\mathbf x,t) + \,\mathbf \nabla\cdot\big( \mathbf v_{\pm}(\mathbf x,t) \rho_{\pm}(\mathbf x,t) \big) = r_{\mp}(\mathbf x,t)\rho_{\mp}(\mathbf x,t) - r_{\pm}(\mathbf x,t)\rho_{\pm}(\mathbf x,t)
	\end{equation}
where,
\begin{gather*}
	\rho_{\pm} = j^{0}_{\pm}= \psi_{\pm}^{\dagger}\psi_{\pm},\quad \mathbf{v}_{\pm} = \frac{\mathbf{j}_{\pm}}{j^{0}_{\pm}} = \frac{\psi_{\pm}^{\dagger}\boldsymbol{\gamma}\psi_{\pm}}{\psi_{\pm}^{\dagger}\psi_{\pm}}\\
	r_{\pm} = \frac{(\pm F)^{+}}{j^{0}_{\pm}} = 2m\frac{\Im\left(\psi_{\pm}^{\dagger}\gamma^{0}\psi_{\mp}\right)^{+}}{\psi_{\mp}^{\dagger}\psi_{\mp}}
\end{gather*}
where $ F^{+} = \max\{F, 0\} $. It was used that $ F = F^{+}-(-F)^{+} $.


Equation \eqref{inr} has exactly the same form as a Master Equation for run-and-tumble particles with a time- and position-dependent tumble rate \cite{tailleur2008statmech}. Such run-and-tumble particles would individually obey the equation of motion
\begin{equation}\label{gui}
	\dot{\mathbf x}_t = \mathbf{v}_{\chi_{{}_{t}}} (\mathbf x_t,t)
\end{equation}
while their densities evolve by \eqref{inr}. We now make a step which is usually not made in the context of quantum physics, but is very natural in statistical mechanics. That is, we will interpret \eqref{inr} exactly as a Master Equation, and accept the Langevin-type dual picture that comes with it. We thus attribute trajectories to the electrons, which are characterized by tumbles induced by the tumbling field $ r_{\pm}(\mathbf{x},t) $. The chirality $ \chi_{{}_{t}} $ of the electrons then follows an inhomogeneous Markov jump process during which it flips between $+$ and $-$, yielding a zig-zag picture of the electron motion; see e.g. \cite[p.\ 632]{pen}.

We emphasize that no approximations have been made to derive \eqref{inr}, and that the motion of the electrons is fully determined by the solution of the Dirac equation $ \psi $.

Given the above interpretation, it is interesting to connect the Cooper pairing mechanism with the emergence of bound states in run-and-tumble particles.
The connection becomes particularly pertinent when we transform eqs. \eqref{eq:eom_model_active} to a (co)moving frame, i.e.,
\begin{equation}\label{eq:moving_frame}
	\begin{aligned}
		\eta_{i} &\to z_{i} = \eta_{i} - \frac{\varepsilon t}{\zeta} \\
		x_{\alpha} &\to y_{\alpha} = x_{\alpha} - \frac{\varepsilon t}{\zeta}
	\end{aligned}
\end{equation}
After which the equations of motion become
\begin{equation}\label{eq:eom_model_moving}
	\begin{aligned}
		\zeta \dot{z}_i(t) &= - \sum_{\alpha=1,2}u'\big(y_{\alpha}(t) - z_i(t)\big) + \left(\frac{2\zeta}{\beta}\right)^{\frac{1}{2}}\xi_i(t)\\
		\gamma \dot{y}_{\alpha}(t) &= \nu_{0}\sigma_{\alpha}(t) - \frac{\gamma}{\zeta}\varepsilon -\sum_{i=1}^Nu'\big(z_i(t)- y_{\alpha}(t)\big)
	\end{aligned}
\end{equation}
where now the colloids at positions $z_i$ are not driven, but are in equilibrium, while the probes at $ y_{\alpha} $ are driven around the ring in the negative-$ y $ direction. The dynamics described by \eqref{eq:eom_model_moving} are exactly equivalent to those described by \eqref{eq:eom_model_active} (see Fig \ref{fig:histogram_passive}), except that the source of the driving \emph{and} the active propulsion is to be found in the probes. 

To summarize, the similarities between Cooper pairs and the model in this paper are: in both cases there are particles driven through a periodic thermal medium in equilibrium, and in both cases the driven particles display run-and-tumble behavior. Finally, both cases display pairing effects at low temperature.

\section{Conclusion}
It is possible to obtain a bound state between active probes in the sense that the histogram of the probe separation shows three peaks.  
The mechanism of the present paper was to confine them to a ring and allow short range interaction with a dilute medium of globally driven colloids.  
The prediction from the quasi-static analysis (where the colloids move much faster than the probes) reveals itself stable against bath fluctuations.  
We observe a transition between the presence and absence of bound states as function of the persistence.  
Going to a co-moving frame, we obtain pair formation between run-and-tumble probes in an external field in a periodic one-dimensional system, when coupled to an equilibrium bath. 
\bibliography{boundst}

\onecolumngrid
\newpage
\appendix
\section{Stationary density profile}\label{sec:coll_profile}
This section provides a summary of results from \cite{maes2017nonreactive} related to the density profile of the driven colloids on a ring with probes. For more detail on the calculations we refer to \cite{maes2017nonreactive}.

The probes are considered to be fixed (quasi-static) at positions $ x_{\alpha} $ where in our case $ \alpha = 1,2 $, but generalizations to larger numbers of probes are possible. There are a large number of driven colloids which are considered to be independent of each other and thus represented by a generic coordinate $ \eta $. The size of the ring is denoted $ L $ and the average density of colloids is $ \rho^{0} $ such that the number of colloids is given by $ \rho^0 L $. The probes and colloids interact through an interaction potential $ u(x_{\alpha}-\eta) $ with properties $ u(z) = u(-z) $ (parity) and $ u(z)=0 $ for $ |z|>\delta $ (locality); the parameter $ \delta $ representing the size of the probes. The dynamics of the colloids may then be represented by the overdamped Langevin equation,
\begin{equation}\label{app:col_dyn}
	\zeta\dot{\eta}(t) = \varepsilon - \frac{\partial}{\partial\eta}U_{x}(\eta) + \left(\frac{2\zeta}{\beta}\right)^{\frac{1}{2}}\xi(t)
\end{equation}
where $ U_{x}(\eta) = \sum_{\alpha=1,2}u(x_{\alpha}-\eta) $ and $ \varepsilon\geq0 $ is a constant driving force. The constants $ \beta $ and $ \zeta $ represent the inverse temperature and friction coefficients respectively. Again, we define a dissipative length scale $ \ldis = (\varepsilon\beta)^{-1} $. The stationary density of the colloids may be found by solving the stationary Smoluchowski equation,
\begin{equation}\label{app:smol}
	0 = -\frac{\partial}{\partial\eta}\left[\left(\varepsilon - \frac{\partial U_x}{\partial\eta}\right)\rho_{x}\right]+\frac{\partial^{2}}{\partial\eta^{2}}\left(\frac{\rho_{x}}{\beta}\right)
\end{equation}
This equation can be solved up to corrections of order $ \mathcal{O}(e^{-L/\ldis}) $,
\begin{equation}\label{app:dens}
	\rho_{x}(\eta) = \frac{\rho^{0}e^{-\beta U_x(\eta)}}{Z(x)}\left[1- \frac{1}{\ldis}\sum_{\alpha}\Phi(x_{\alpha}-\eta)e^{-(x_{\alpha}-\eta)/\ldis}\right]
\end{equation}
Where we have introduced $ \Phi(z) = \int_{-\delta}^{z}\dif z'\phi^{+}(z')e^{z'/\ldis} $ with $ \phi^{\pm}(z) = \pm1\mp e^{\pm\beta u(z)} $; the normalization is given by
\begin{equation}
	Z(x) = 1 + \frac{2A}{L} - \frac{B}{L\ldis}\sum_{\alpha, \gamma\neq\alpha}e^{-(x_{\alpha}-x_{\gamma})^{+}/\ldis}
\end{equation}
with the notation $ z^{+} $ introduced in \eqref{eq:onewaysep}. We have now also introduced the form factors
\begin{equation*}
	\begin{aligned}
		A &= \int_{-\delta}^{\delta}\phi^{-}\phi^{+}(z)\,\dif z -\frac{1}{\ldis}\int_{-\delta}^{\delta}\dif z\int_{-\delta}^{z}\dif z'\,
\phi^{-}(z)\phi^{+}(z')\,e^{(z'-z)/\ldis}\\
		B &= B^{+}B^{-},\quad\text{with}\ B^{\pm} = \int_{-\delta}^{\delta}\phi^{\pm}(z)\, e^{z/\ldis}\, \dif z
	\end{aligned}
\end{equation*}
which encode the properties of the probe-colloid interaction.

From \eqref{app:dens}, we find the stationary colloidal current,
\begin{equation}\label{app:current}
	j_{x} = \frac{\varepsilon\rho^{0}}{\zeta Z(x)}
\end{equation}
which is bounded from above by the free colloidal current without probes $ j_{x} \leq \varepsilon\rho^{0}/\zeta $ which constrains $ Z(x)\geq1 $.

Using \eqref{app:dens} and \eqref{eq:avg_force} allows us to quantify $ f^{\mathrm{drift}} $ and $ f^{\mathrm{int}}_{\alpha} $ as
\begin{equation}
	\begin{aligned}
		f^{\mathrm{drift}} &= \zeta A j_{x}\\
		f^{\mathrm{int}}_{\alpha} &= -\frac{\zeta j_{x}B}{\ldis}\sum_{\gamma\neq\alpha}e^{-(x_{\gamma}-x_{\alpha})^{+}/\ldis}
	\end{aligned}
\end{equation}
which allows for the identification of $ \kappa $ in \eqref{eq:dyn_d} as
\begin{equation}\label{eq:kappa}
	\kappa = \frac{\zeta j_{x}B}{\ldis}.
\end{equation}

\section{Solution to $ \gamma\dot{d} = F(d) + \nu$}\label{sec:sol_d}
The quasi-static approximation \eqref{eq:eom_d_active} for the probe separation $d = d(t)$ gives
\begin{equation}\label{eq:app_A1}
\gamma \dot{d}(t) = \kappa\left(e^{\frac{-d(t)}{\ldis}}-e^{\frac{d(t)-L}{\ldis}}\right) + \nu
\end{equation}
representing the dynamics of the separation between two probes which interact through one-sided exponential tails.  It is easily solved from
\begin{equation}\label{eq:sep_var}
\int_{d_{0}}^d\frac{\dif x}{\kappa\left(e^{-\frac{x}{\ldis}}-\Lambda e^{\frac{x}{\ldis}}\right)+\nu} = \frac{1}{\gamma}\, t
\end{equation}
where  $ \Lambda = e^{-\frac{L}{\ldis}} $. Solving the integral yields
\begin{equation*}
\int_{d_{0}}^d\frac{\dif x}{\kappa\left(e^{-\frac{x}{\ldis}}-\Lambda e^{\frac{x}{\ldis}}\right)+\nu} = \frac{\ldis}{\mu}\log\left|\frac{2\kappa e^{-\frac{d}{\ldis}}+\nu+\mu}{2\kappa e^{-\frac{d}{\ldis}}+\nu-\mu}\right| + C_0,\quad\mu=\sqrt{4\kappa^{2}\Lambda+\nu^{2}}
\end{equation*}
where the constant $C_0$ accounts for the initial condition. From \eqref{eq:sep_var}, 
\begin{equation*}
\left|\frac{2\kappa e^{-\frac{d}{\ldis}}+\nu+\mu}{2\kappa e^{-\frac{d}{\ldis}}+\nu-\mu}\right| = \left|\frac{2\kappa e^{-\frac{d_{0}}{\ldis}}+\nu+\mu}{2\kappa e^{-\frac{d_{0}}{\ldis}}+\nu-\mu}\right|e^{\frac{\mu}{\gamma\ldis}t}= C(t)
\end{equation*}
where $d_0 = d(t=0)$.  Solving for $d(t)$, we get the expression
\begin{equation}
d(t) = -\ldis\log\left[-\frac{\mu}{2\kappa}\left(\frac{1+C(t)}{1-C(t)}\right)-\frac{\nu}{2\kappa}\right]
\end{equation}

In the limit  $ t\to\infty $, the solution $d(t)$ tends to $ -\ldis\log\frac{\mu-\nu}{2\kappa} $, which is the fixed point of \eqref{eq:app_A1}. When $\nu=0$, the results from (\ref{eq:dyn_d}, \ref{eq:sol_d}) are recovered, and the solution tends to $ L/2 $.

\section{Simulation details}\label{app:simulations}
The simulations were performed using a forward Euler-type Brownian dynamics simulation \cite{algorithms}, the discretized equations of motion being given by
\begin{equation}\label{eq:eom_sim}
\begin{aligned}
\eta_i(\Delta t) &= \eta_{i,0} + \frac{1}{\zeta}\left(\epsilon - \sum_{\alpha}u'(x_{\alpha,0} - \eta_{i,0})\right)\Delta t + \left(\frac{2\Delta t}{\beta\zeta}\right)^{\frac{1}{2}}\Gamma\\
x_{\alpha}(\Delta t) &= x_{\alpha, 0} + \frac{1}{\gamma}\left(-\sum_{i}u'(\eta_{i,0}-x_{\alpha,0})\right)\Delta t
\end{aligned}
\end{equation}
where $ \Delta t $ represents one timestep, and the quantities with subscript \textsubscript{0} are those at the beginning of the timestep. 
$\Gamma$ is a Gaussian random variable with zero mean and unit variance. 

In addition to the parameters listed in Tab. \ref{tab:params}, $ \Delta t $ was usually set to $ \Delta t = 10^{-3} $.

A working code example can be found in \cite{gitlab}.

\end{document}